# Application and Assessment of Deep Learning for the Generation of Potential NMDA Receptor Antagonists

*Katherine J. Schultz, Sean M. Colby, Yasemin Yesiltepe, Jamie R. Nuñez, Monee Y. McGrady, Ryan R. Renslow\**

Pacific Northwest National Laboratory, Richland, WA, USA.

ABSTRACT. Uncompetitive antagonists of the *N*-methyl D-aspartate receptor (NMDAR) have demonstrated therapeutic benefit in the treatment of neurological diseases such as Parkinson's and Alzheimer's, but some also cause dissociative effects that have led to the synthesis of illicit drugs. The ability to generate NMDAR antagonists *in silico* is therefore desirable both for new medication development and for preempting and identifying new designer drugs. Recently, generative deep learning models have been applied to *de novo* drug design as a means to expand the amount of chemical space that can be explored for potential drug-like compounds. In this study, we assess the application of a generative model to the NMDAR to achieve two primary objectives: (i) the creation and release of a comprehensive library of experimentally validated NMDAR phencyclidine (PCP) site antagonists to assist the drug discovery community and (ii) an analysis of both the advantages conferred by applying such generative artificial intelligence models to drug design and the current limitations of the approach. We apply, and provide source code for, a variety of ligand- and structure-based assessment techniques used in standard drug discovery analyses to the deep learning-generated compounds. We present twelve candidate antagonists that are not available in existing chemical databases to provide an example of what this type of workflow can achieve, though synthesis and experimental validation of these compounds is still required.

INTRODUCTION. Deep learning (DL)-based technology has been widely adopted in realms such as natural language processing and computer vision. The success of DL frameworks in those arenas led to their more recent adaption to the cheminformatics field, where they have been employed as tools to further elucidate topics ranging from chemical property prediction to synthesis planning.[1-2] Furthermore, DL is playing an increasingly important role in the *de novo* design of molecules with desired chemical properties. It is estimated that there are up to $10^{60}$ synthesizable drug-like organic compounds in chemical space, over 99.9% of which are theoretically accessible but have never been synthesized.[3-4] The *in silico* design and property prediction of biologically active compounds is desirable both as a means to explore previously inaccessible areas of chemical space and as a labor-, cost-, and time-effective alternative to the *in vitro* assessment of lead-like compounds via high throughput screening. As such, there has been a sizable improvement to *in silico* methodologies for novel molecule generation, discovery, and property prediction in the past decade. These computer-assisted drug discovery (CADD) methods also have the potential to achieve much higher hit-finding rates than laboratory-based high-throughput screening.[5] While the application of machine learning to the CADD pipeline in the areas of target, property, and activity prediction is widespread,[6-7] its application to *de novo* molecular generation is a more recent development—one that, while not currently without notable limitations, is demonstrating great potential.

One standard *in silico* approach to *de novo* molecular design is to build a molecule piecewise or atom-wise inside a pocket that represents the target protein.[8] However, this method can result in molecules that are difficult to synthesize, overfit to their target, or both. Another approach is to use virtual chemical reactions to build novel molecules. While this method addresses the synthesizability problem, it limits the scope of chemical space that can be explored.[1, 9] To avoid these pitfalls, inverse quantitative structure-activity relationships (inverse-QSAR) has been proposed as an alternate approach to *de novo* molecular design. The methodology employed by inverse-QSAR is to generate compounds by sampling the region of chemical space defined by the properties of molecules with known activity.[10-12] To this end, DL can be utilized both for learning the property space encompassed by the active compounds and for generating novel molecules from this space.[9, 13-21] A promising recent development involves the application of variational autoencoders (VAEs)[22] to *de novo* molecular design.[14, 23-25] VAEs are generative DL models comprised of two connected networks—an encoder and a decoder. The encoder converts input data to a compact representation, shaping the network's continuous – or latent – space as it learns patterns in the input. The decoder samples from the latent space to generate the output. In the case of *de novo* design, the VAE's latent space represents a chemical space within which known molecules can be placed and from which novel molecular structures can be derived. VAE architecture is particularly compelling for inverse-QSAR due to its ability to accommodate two-way traversal: chemical property prediction given input molecular structure or structure generation from desired properties.

Recently, our group developed the VAE-based software DarkChem for mapping chemical properties to molecular structure.[26] It was originally intended as a tool for *in silico* metabolomics to aid in the identification of small molecules in complex samples. DarkChem's latent space is shaped using calculable chemical properties, such as molecular mass and collision cross section (a gas phase property measured by ion mobility spectrometry), to enable rapid construction of massive molecular libraries with reduced reliance on experimentation. DarkChem also shows promise in *de novo* drug design. When a small set of known channel blockers of the NMDAR (*N*-methyl D-aspartate receptor) were used as input to DarkChem to generate putative novel compounds, principal component analysis (PCA) revealed that the DarkChem-generated molecules clustered well according to mass and CCS and warranted a more thorough investigation of their capacity as potential NMDAR channel blockers. In an effort to further elucidate both the advantages and shortcomings of nascent artificial intelligence-assisted generative approaches like DarkChem, we chose to enter the field of computational drug design by exploring how our computer-generated molecules perform when applied to filtering and assessment methods commonly used during virtual screening of existing large molecular databases. It should be noted that while synthesizability was assessed during the course of this study, none of the computer-generated molecules have been synthesized, as this is outside the scope of our work. Our findings are intended to offer guidance both to the development of future versions of DarkChem and to the drug design community. We opted to conduct our exploration on a target that has received renewed interest for drug leads of late,[27-29] yet is lacking in readily available and accessible information necessary for many CADD techniques: the NMDAR phencyclidine (PCP) site. As such, a secondary goal of this study was to provide a comprehensive PCP site library freely to the research community. Furthermore, we evaluated our generative artificial intelligence (AI) model against a more challenging target – an active site located in the ion channel of a complex plasma membrane protein only recently resolved by X-ray crystallography[30] – than those used in literature to date, which are largely kinases.[16, 18, 25, 31-32]

The NMDAR PCP Site

The NMDAR is a heterotetrameric, ligand- and voltage-gated glutamate receptor expressed in the central nervous system (CNS). Irregular NMDAR function is implicated in excitotoxicity and a host of nervous system disorders including Alzheimer's, Parkinson's, and Huntington's diseases. A connection between NMDAR dysfunction and depression has also been proposed, but experiments aimed at elucidating the mode of action have yielded ambiguous results.[33-34] While there are multiple binding sites on the NMDAR (Figure 1), open channel blockers that target the PCP site in the ion channel of the transmembrane domain (TMD) have received renewed interest due to the recent clinical trial successes of ketamine (a PCP site uncompetitive antagonist) in treating major depressive disorder.[27-29] Memantine is a PCP site antagonist that is well-tolerated and effective in the treatment of Parkinson's and moderate to severe Alzheimer's disease.[35] Many other PCP site antagonists, including the site's namesake drug PCP and the extremely high-affinity channel blocker MK-801 (dizocilpine), result in undesirable dissociative effects in humans that negate the potential therapeutic benefit they might confer. Further, many PCP analogs have been identified in confiscated street drugs and in the post-mortem setting.[36] Due to the nature of known PCP site antagonists, the potential benefit of generating novel analogous structures is twofold: to aid in the discovery of medicinal drug leads for the treatment of neurological disorders and to preempt new designer drugs before they hit the market.

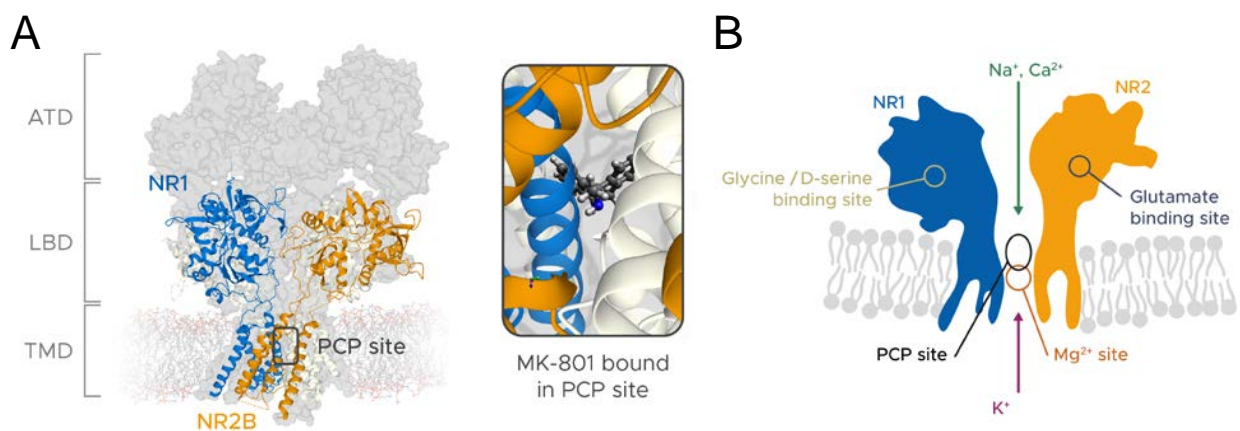

**Figure 1.** The NMDAR and some of its notable features. (A) The crystal structure of the NMDAR derived from Protein Data Bank accession no. 5UOW (Lu et al., 2017) with the amino-terminal domain (ATD), ligand-binding domain (LBD), and transmembrane domain (TMD) regions

labeled. NR1 and NR2B subunits are isolated for clarity. The high-affinity channel blocker MK-801 is shown bound in the PCP site in the inset. (B) Cation flow and selected binding sites including the glycine and glutamate agonist binding sites are depicted in the schematic.

Despite the significant value the NMDAR has as a target, DL-enabled *de novo* drug design has not been applied to the NMDAR to date – in part due to the lack of publicly available ligand-protein binding data. To address this data scarcity, we have built a comprehensive NMDAR PCP site antagonist library (see Supporting Information). Another impediment to CADD application to the NMDAR – and particularly to the PCP site – has been the difficulty in resolving the protein in its active state via X-ray crystallography due to its complex structure.[37] In 2018, however, Song et al. succeeded in resolving a PCP site-bound crystal structure of the NMDAR with a correctly assembled TMD channel,[30] an achievement noted in part for providing a good assessment tool for the binding of NMDAR channel blockers.[37] We therefore elected to include a docking study of PCP site antagonists using this structure as a component of our assessment.

Given the high failure rate of NMDAR antagonists in clinical trials and the frequency with which new street drugs that target the PCP site are developed, a primary goal of the molecular generation aspect of this study was to produce compounds that are structurally unique compared to known PCP site antagonists yet still retain predicted target activity. The discovery of new molecular scaffolds and chemical classes with PCP site activity has the potential to aid the development of therapeutics without undesirable dissociative effects. In our effort to explore the efficacy and limitation of employing AI for molecular generation while providing comprehensive information about PCP site antagonists heretofore missing from existing publicly available small molecule databases, we utilized a variety of established *in silico* techniques currently being used to find candidate drug leads. These include ligand- and structure-based methods for activity assessment, absorption, distribution, metabolism, and excretion (ADME) prediction, substructure analysis, lead-likeness filters, synthesizability scoring, and similarity metrics. Ultimately, we found twelve new potential antagonists that passed all of our filtering steps. All structures were not present in any public database and contained unique molecular backbones compared to known active antagonists. The results of applying these techniques to AI-generated compounds provide insight into some of the advantages afforded by DL for *de novo* drug design, as well as the obstacles. The limitations we identified have the potential to be useful in guiding both our future work and, more broadly, the development of generative machine learning models for targeted molecule design.

RESULTS AND DISCUSSION.

*In silico* Workflow Strategy

One of our primary objectives was to assess whether our AI-generated compounds would pass *in silico* screens typically utilized during the CADD process to characterize and filter existing compounds or compounds created by other *in silico* or *in vitro* means. We therefore developed a workflow comprised of both ligand- and structure-based techniques commonly used in the drug discovery process to screen AI-generated compounds for potential activity at the NMDAR PCP site (Figure 2). Our strategy for the creation of *de novo* candidate PCP site antagonists was to search the chemical space (i.e. the VAE latent space) encompassed by the set of experimentally verified PCP site antagonists. This necessitated the creation of a library of known actives to define a region in latent space from which to sample. Resulting generated compounds were filtered according to predicted activity at the binding site, favorable molecular docking score, and desirable ADME-Toxicity profiles and synthesizability scores.

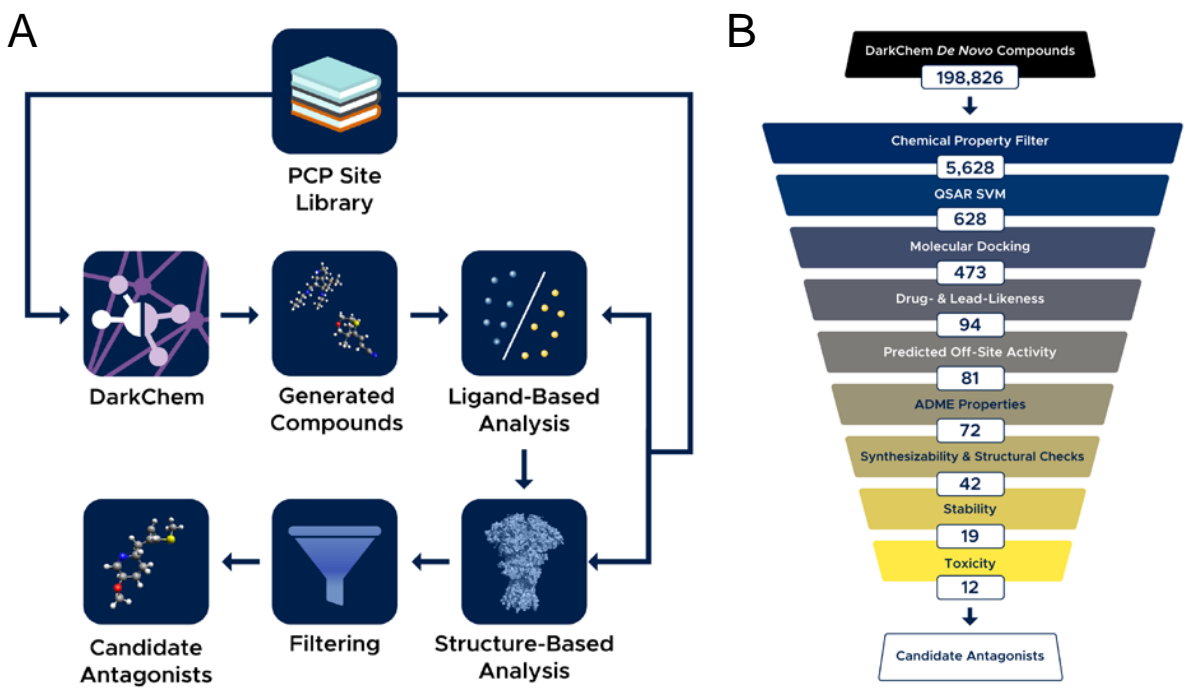

**Figure 2.** (A) Workflow schematic: A PCP site library was assembled and used as input to DarkChem to train an activity prediction model and collect baseline performance metrics for molecular docking simulations. Generated candidate structures were assessed with the ligand-based activity prediction model and the structure-based molecular docking simulation and then further filtered. (B) Detailed representation of the filtering process for AI-generated compounds with the number of passing compounds displayed below the filtering step. All molecules that pass each step are available upon request.

NMDAR PCP Site Molecular Library

The lack of publicly available comprehensive data on NMDAR channel blockers necessitated the construction of a library of NMDAR PCP site antagonists. We created a library that contains 1,142 NMDAR PCP site antagonists, with 818 compounds having experimentally derived activity at the PCP site and 324 compounds having unknown activity. Some of the dominant chemical classes represented include arylcyclohexylamine, dibenzocycloalkenamine, dioxolane, benzomorphan, diphenylethylamine, aminoadamantane, aminoalkylcyclohexane, morphinan, and guanidine. Classes are primarily defined according to the classifications used by the author(s) in the publication from which the compounds were extracted. If such classifications were not provided, applicable class names already present in the library were applied. The library was built from an in-depth literature search and web scraping of online databases. A summary of the literature-derived library sources is shown in Table 1. Many well-known NMDAR antagonists and their analogs are represented, such as PCP, rolicyclidine, ketamine and esketamine, tiletamine, lanicemine, methoxetamine, dexoxadrol and etoxadrol, memantine, dextromethorphan and dextrorphan, and dizocilpine (MK-801). The chemical properties of the library entries reveal some common trends among them including a low molecular weight (μ: 261.03 Da; σ: 63.79 Da) and high lipophilicity (μ: 3.75; σ: 1.29) compared to other CNS-active drugs.[38] The library also contains 2,000 decoy molecules—compounds with physicochemical properties that match actives but are topologically different and presumed to be inactive at the site of interest. All compounds are represented by their International Union of Pure and Applied Chemistry (IUPAC) name, common name where applicable, canonical simplified molecular-input line-entry specification (SMILES), and chemical formula. The library was utilized to define a region in the DarkChem latent space to search for novel antagonists, build an activity prediction model during ligand-based assessment, and determine baseline molecular docking performance metrics.

**Table 1. Summary of NMDAR PCP site library sources.** The full library spreadsheet is available in the Supporting Information.

| Chemical Class(es) | Source | Citation | Number of Compounds* |
|---|---|---|---|
| Dioxolane | Aepkers & Wünsch, 2005 | [39] | 5 |
| Aminoacridine | Barygin et al., 2009 | [40] | 5 |
| Arylcyclohexylamine, Diphenylethylamine | Berger et al., 1998, 2009, 2015 | [41-43] | 36 |
| Arylcyclohexylamine, Dibenzocycloalkenamine, Octahydrophenanthrenamine | Bigge et al., 1993 | [44] | 47 |
| Arylcyclohexylamine | Chaudieu et al., 1989 | [45] | 33 |
| Arylcyclohexylamine, Arylcyclohexylmorpholine | Colestock et al., 2018 | [46] | 7 |
| Arylcyclohexylamine, Benzomorphan, Indenopyridine, Benz(f)isoquinoline, Dioxolane | Domino & Kamenka, 1988 | [47] | 43 |
| Arylcyclohexylamine, Guanidine | Dravid et al., 2007 | [48] | 7 |
| Phenylpiperidine | Ebert et al., 1998 | [49] | 18 |
| Dibenzocycloalkenamine | Elhallaoui et al., 2003 | [50] | 36 |
| Dibenzocycloalkenamine | Gee et al., 1993, 1994 | [51,52] | 24 |
| Aminoalkylcyclohexane | Gilling et al., 2007 | [53] | 3 |
| Quinolone | Gordon et al., 2001 | [54] | 1 |
| Tetrahydroisoquinoline, Benzomorphan | Gray et al., 1989 | [55] | 38 |
| Hexahydrofluorenamine | Hays et al., 1993 | [56] | 22 |
| Guanidine | Hu et al., 1997 | [57] | 66 |
| Arylcyclohexylamine | Itzhak et al., 1981 | [58] | 2 |
| Diphenylethylamine | Kang et al., 2017 | [59] | 1 |
| Arylcyclohexylamine | Kozikowski & Pang, 1990 | [60] | 12 |
| Arylcyclohexylamine, Propanolamine | Kozlowski et al., 1986 | [61] | 9 |
| Arylcyclohexylamine, Dioxolane, Benzomorphan | Largent et al., 1986 | [62] | 6 |
| Arylcyclohexylamine, Isothiocyanate | Linders et al., 1993 | [63] | 25 |
| Arylcyclohexylamine, Dioxolane, Benzomorphan, Benz(f)isoquinoline, Morphinan | Mendelsohn et al., 1984 | [64] | 15 |
| Dibenzocycloalkenamine | Monn et al., 1990 | [65] | 20 |
| Arylmethylguanidine | Naumiec et al., 2015 | [66] | 17 |
| Aminoalkylcyclohexylamine, Tetrahydroisoquinoline, Imidazoline | Nicholson & Balster, 2003 | [67] | 10 |
| Arylcyclohexylamine, Imidazoline | Olmos et al., 1996 | [68] | 9 |
| Aminoalkylcyclohexane, Aminoadamantane | Parsons et al., 1995, 1999 | [69, 70] | 55 |
| Arylcyclohexylamine, Adamantine, Aminoadamantane, Aminoalkylcyclohexane | Rammes et al., 2001 | [71] | 15 |
| Dibenzocycloalkenamine | Rogawski et al., 1991 | [72] | 1 |
| Arylcyclohexylamine, Anisylcyclohexylamine | Roth et al., 2013 | [73] | 6 |
| Arylcyclohexylamine | Sałat et al., 2015 | [74] | 2 |
| Dioxolane | Sax & Wünsch, 2006; Sax et al., 2008 | [75, 76] | 71 |
| Arylcyclohexylamine | Stefek et al., 1990 | [77] | 4 |
| Diphenylethylamine | Subramaniam et al., 1996 | [78] | 2 |
| Dibenzocycloalkenamine | Thompson et al., 1990 | [79] | 73 |
| Arylcyclohexylamine, Arylcycloheptylamine | Thurkauf et al., 1990 | [80] | 37 |
| Aminoadamantane | Tikhonova et al., 2004 | [81] | 14 |
| Arylcyclohexylamine, Anisylcyclohexylamine, Diphenylethylamine, Benzofuran, Dioxolane | Wallach, 2014; Wallach et al., 2016; Wallach & Brandt, 2018 | [36, 82-85] | 77 |
| Morphinan | Werling et al., 2007 | [86] | 2 |
| Arylcyclohexylamine, Anisylcyclohexylamine | Zarantonello et al., 2011 | [87] | 12 |
| Arylcyclohexylamine | Zukin & Zukin, 1979 | [88] | 31 |

*some compounds appear in multiple sources

Generation of Potential PCP Site Antagonists

   Our DL-based VAE, DarkChem, was supplied with known actives from the library to seed its latent representation of chemical structure, or latent space, defining a bounded volume within chemical space from which to sample new potential NMDAR antagonists (see Figure 3 and Methods). Ensuing deduplication and SMILES validity verification steps resulted in a set of 198,826 generated structures. Among this set, all required chemical properties for the activity model could be successfully calculated for 5,629 compounds. The remaining downselection steps, which are detailed in following sections, further reduced the candidates to twelve optimized potential antagonists (Figure 4), none of which were found in the thirty-one chemical databases we cross-referenced. These final putative structures were assessed for their proximity to the set of known actives to determine whether generated structures were unique and/or novel, as opposed to simple perturbations of the input. The distance between each putative structure and active in the 128-dimension latent space was computed by L1 norm, as it behaves favorably in high dimensions compared to the L2 norm,[89] and the closest active was selected for each generated structure (Table S5). Example final structures are shown with their closest therapeutic active in latent space in Figure 3. The primary drawback to the *in silico* generation of novel molecules is the unavailability of chemical standards for an *in vitro* assessment of their lead-likeness. When (and if) these *de novo* compounds are synthesized, their ability as PCP site antagonists can be assessed to improve the accuracy of the *in silico* pipeline where needed.

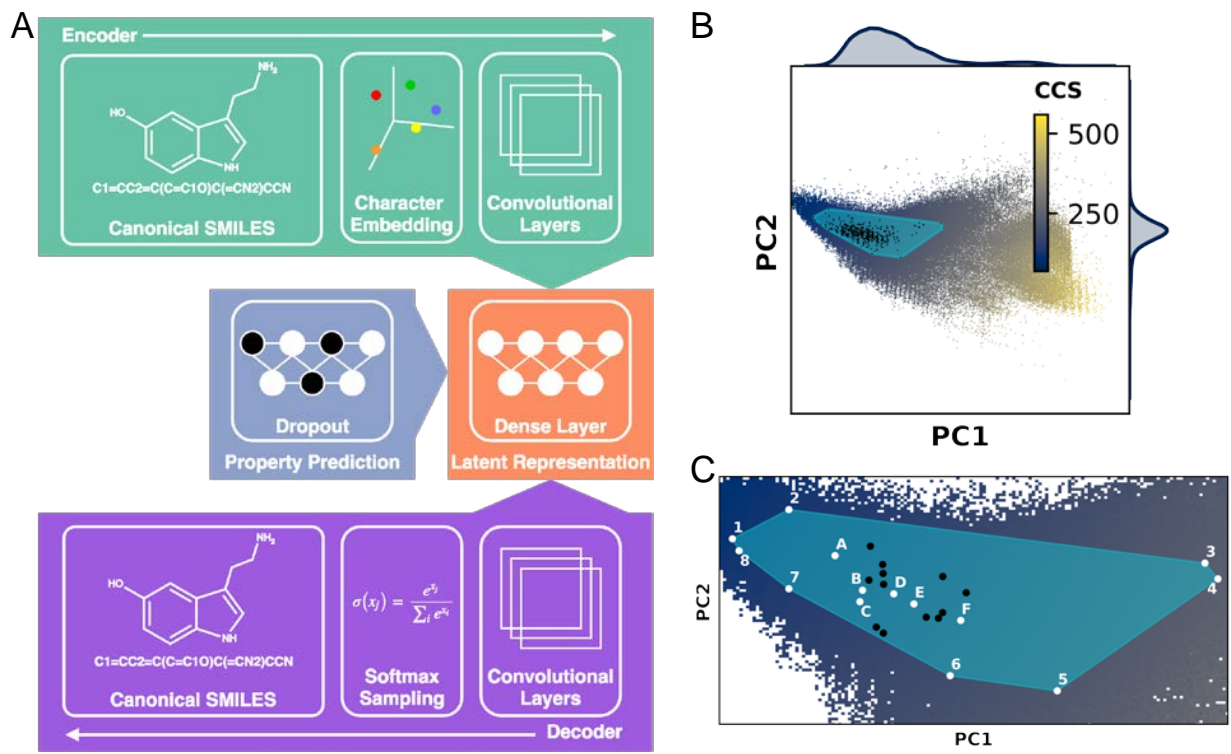

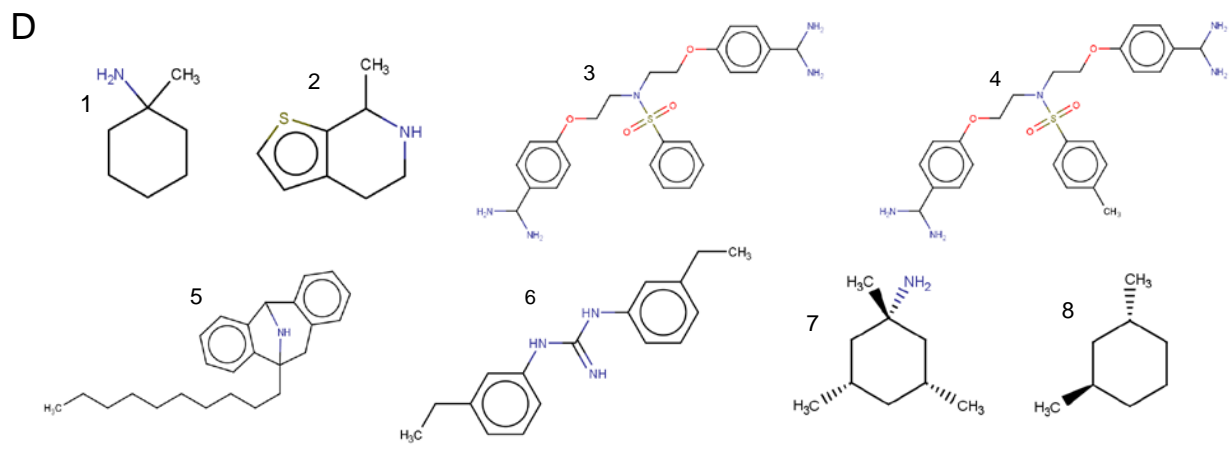

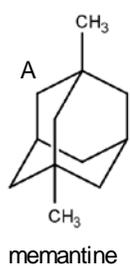 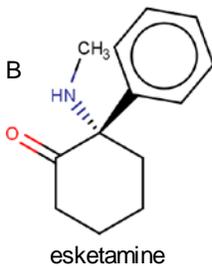 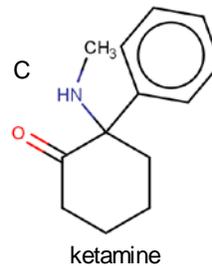

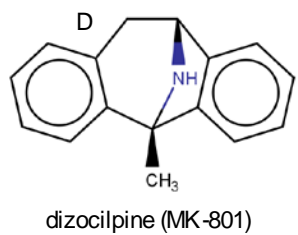 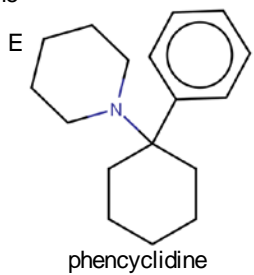 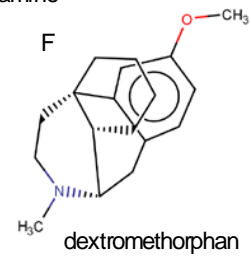
12

**Figure 3.** (A) DarkChem network schematic: The network involved an encoder (green), a latent representation (orange), and a decoder (purple). Additionally, a property predictor (slate) was attached to the latent representation. For the encoder, layers included SMILES input, character embedding, and a series of convolutional layers. The latent representation was a fully connected dense layer. The decoder was comprised of convolutional layers, followed by a linear layer with softmax activation to yield outputs. Finally, the property predictor was a single dense layer connected to the latent representation with 20% dropout. Reprinted with permission from Colby et al., 2019. Copyright 2019 American Chemical Society. (B) Latent space: The first two principal components of the 128-dimensional representation are shown, colored by predicted property value (top: m/z, bottom: CCS). The representation is a 2D binned statistic of the mean, with grid size 384 in each principal component dimension. A kernel density estimator is also shown for each principal component dimension, emphasizing density of the distribution. In teal is the region of latent space encompassed by PCP site antagonists. (C) Close-up of PCP site antagonist region of latent space: Points *1-8* represent a selection of known PCP site antagonists, and the therapeutic subset of known antagonists is represented as *A-F*. (D) Molecular structures corresponding to points *1-8* and *A-F*.

Ligand-Based Analysis

Drug screening is a costly and time-consuming process. Many attempts have been made to expedite the process by developing metrics to filter lead-like compounds from large libraries of contenders based on the physiocochemical properties of the ligands alone. The most famous of these metrics is arguably Lipinski's Rule of Five (Ro5).[90] There are many exceptions to the Ro5, however, and numerous other metrics have been developed in an effort to improve upon it. The current consensus regarding these rules is that while they are a good starting place for filtering candidate compounds, they are far from comprehensive. In addition, the Ro5 and similar rules have been demonstrated to be less applicable to CNS active compounds.[91] They are, however, approachable and easy to implement.

In our work, once properties were calculated for PCP site library compounds, the Ro5 and several other filters were applied to ascertain whether they could effectively distinguish between active and inactive/decoy compounds. None of the filters performed meaningfully better on the actives, and in fact many filtered out a higher proportion of actives than inactives and decoys. This is likely due in part to the highly tuned nature of the inactives and decoys in the library and lends further support to the maxim that such rules should only be applied as an early step when filtering very large and chemically diverse compound databases. Thus, we applied a more rigorous ligand-based analysis technique, detailed below, which resulted in a substantially enhanced ability to distinguish between active and inactive library compounds.

Initially, 130 0-, 1-, and 2-D properties, including atom and bond counts, functional group counts, and topological indices, respectively, were calculated for all library compounds for property analysis. We omitted 3-D properties because they greatly increase computational resources and often do not achieve superior performance to 2-D QSAR methods.[92-94] Applying the library compounds to the principal component analysis (PCA) space of computed properties revealed that the actives tend to cluster tightly while the inactives routinely display much larger variance (Figure S1). Calculated chemical properties were applied to supervised machine learning activity analysis to build a QSAR model. The number of descriptors used in building the model was chosen to balance the "curse of dimensionality", wherein too many descriptors result in decreased model performance, with the loss of relevant information that can occur with too few features. This approach also follows the recommendation that for building machine learning models for QSAR analysis the number of instances should be at least five times the number of features.[95]

Although developed fairly recently, support vector machines (SVM) have demonstrated significant value as activity prediction models.[96-97] The SVM model for PCP site activity prediction achieved a ten-fold cross-validated accuracy of 0.95, with a weighted average precision of 0.98, recall of 0.97, and f1 of 0.97 on the test set (Figure S2). We built the initial classification models using a training set comprised of samples from only the library actives and inactives, which resulted in worse than desired performance due to the scarcity of inactives (N=X). Incorporation of decoy compounds (N=Y) to the inactive class resulted in the considerably more accurate and robust final model built for activity prediction. Decoys are an integral part of the CADD process, used both as controls for assessing docking simulation results and for building QSAR models in the absence of sufficient known inactive compounds. While there is a degree of inherent uncertainty when using decoys, as presumed inactivity does not necessarily equate to true inactivity, the application of an unpaired t-test on the binding outcomes of a docking study of library actives, inactives, and decoys demonstrated statistically significant differences between actives and inactives and between actives and decoys that did not exist between inactives and decoys. This is discussed in further detail in the next section. These results provided confidence in incorporating the decoys to the inactive class for activity model training. To avoid overfitting, a frequent pitfall of QSAR models,[98] we assessed the robustness of our model using both ten-fold cross-validation and a randomly generated test set comprised of data not seen in training. The model's cross-validated accuracy of 0.95 and AUPR of 0.95 demonstrate its usefulness as an assessment tool for PCP site activity. The code for the SVM is available as a Python notebook in the Supporting Information. The SVM model predicted 628 of the 5,628 generated compounds with successfully calculated properties as active.

Structure-Based Analysis

There are two 3D crystallographic NMDAR structures with a PCP site-bound ligand available: Protein Data Bank (PDB) codes 5UOW[99] and 5UN1.[30] The structures were compared by running trial docking simulations following the same receptor preparation method, after which the higher resolution structure 5UN1 (3.6 Å) was chosen for the library and *de novo* compound docking studies. AutoDock Vina outputs the top binding poses and scores for each ligand. A ligand's score represents the estimated free energy of binding, where a more negative score corresponds to a higher likelihood of binding. Vina's energy of binding estimates contain widely reported inconsistencies and should not be assumed to represent correct values for the purposes of ranking, but they have been demonstrated to be accurate in predicting binding poses and distinguishing between active and inactive compounds in the aggregate.[100] A common method for docking assessment is to determine how well docking scores correlate to experimental binding affinities for a given library of compounds. However, this approach is problematic due to the fact that experimental binding affinities present in ligand libraries are extracted from multiple sources with varying experimental parameters, including the location of the receptor in the body. We therefore decided on an alternative approach to docking wherein the ability to distinguish between verified active, verified inactive, and decoy compounds using docking scores was assessed. We found that the docking scores for the library compounds display a statistically significant difference between the active and inactive compounds as well as between the active and decoy compounds (two-tailed $p < 0.0001$ for both), but not between the decoys and inactives (two-tailed $p = 0.3$, Table S2). This result lends support to the use of docking studies for assessing potential activity of novel PCP site antagonists and, more generally, provides confidence in this approach to docking for activity analysis.

A noteworthy finding was that the mean docking score for the DarkChem-generated compounds was less negative than for any of the library compound classes. While individual docking scores were not found to correlate with activity, the mean scores for each library class were distinguishable. This finding could reflect a shortcoming of the present generative AI approach and indicate a need to incorporate receptor structural data into the molecular generation process.

A comparative assessment of generative models with and without the inclusion of structural information would be a valuable area for future work, and we are currently expanding DarkChem to include hundreds of chemical properties along with a valid-structure discriminator in order to improve candidate generation.

A common bottleneck in DL models is the requirement of large quantities of high-quality data. In drug discovery DL applications as in others, more information is typically correlated with better model performance.[101] Therefore, for this work we selected the 473 molecules with docking scores in the actives scoring range to proceed to the next stages of filtering.

Further Assessment of Generated Compounds and Top Final Leads

None of the generated structures were found to exist in the thirty-three compound databases that were cross-referenced for duplicates. ADME-Toxicity, predicted off-site activity, synthesizability, and stability filters were applied to the remaining compounds next (details in Methods). In line with the goal of assisting in the identification of unique scaffolds of molecules with PCP site activity to aid in the search for potentially therapeutic compounds, the filter tolerances were set to be consistent with the properties of known therapeutic actives from the library. This screening process reduced the number of compounds to twelve finalists (Figure 4). As the failure to accurately account for breakdown products has been implicated as the cause of many lead failures during the drug discovery process,[102] we predicted metabolites for each of the twelve candidates (see metabolites xlsx file in Supporting Information).

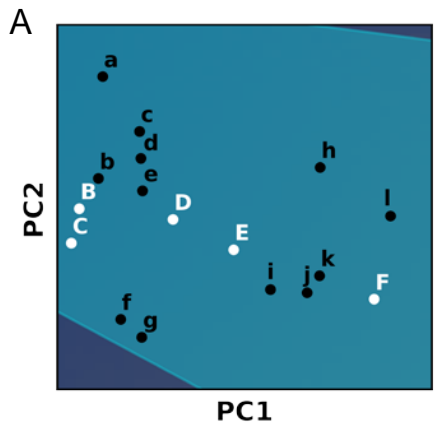

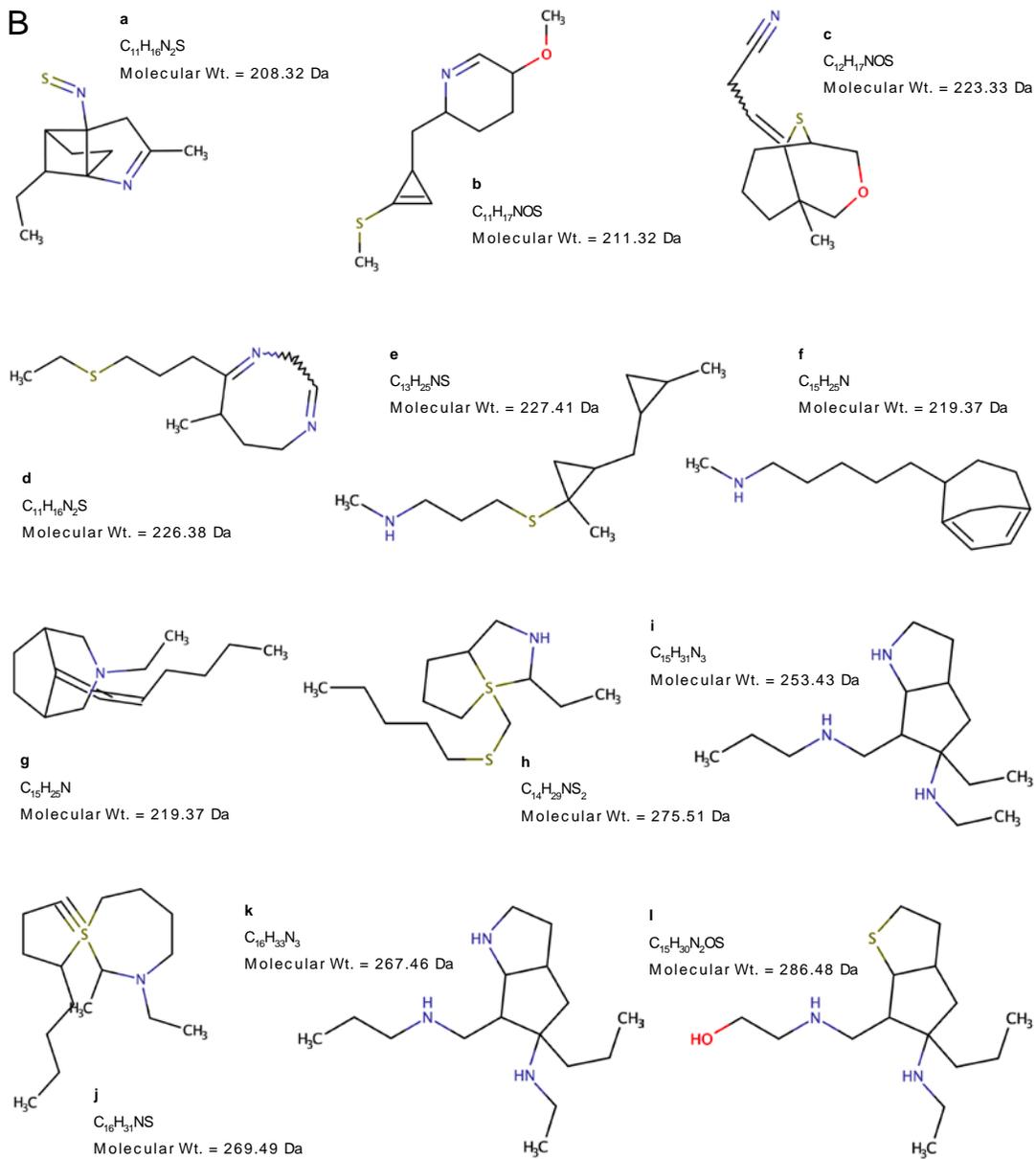

**Figure 4.** (A) Close-up of the first two PCA dimensions from DarkChem's latent space encompassed by the twelve generated finalist molecules, labeled as *a-l*. (B) The twelve generated finalist structures with their letter designation corresponding to that in A.

Similarity assessment of the final structures to the library actives training set revealed that the generated compounds were unique and not merely simple perturbations of the input. Of the twenty most common unique substructures found in at least 50% of the library actives, each substructure is represented in no more than 0.25% of the set of generated compounds. In addition, each of the final twelve compounds contain only between one and eight of the twenty common substructures (Table S3, Table S4). Finalist compounds also display a high degree of uniqueness when compared to the training set using L1 distance and Tanimoto score metrics (Figure 5). The most similar active training set compounds by L1 distance in 128-dimensional space and by Tanimoto score are reported for each finalist molecule in Table S5. The ability of generated compounds to pass multiple stages of standard CADD filtering processes demonstrates the promise of this technique in application to challenging targets. However, the lack of *in vitro* and *in vivo* experimental data for verification and benchmarks for quality assessment pose immediate challenges to a robust assessment of this and other DL-assisted generative models. Furthermore, while the finalist molecules all have favorable synthesizability scores, true synthesizability and stability are difficult to assess and many of the generated compounds are somewhat peculiar looking. To improve DL-assisted *de novo* design methods, representing compounds for DL models as molecular graph convolutions instead of SMILES strings has shown promise as a superior mode of structural representation.[103] We plan to transition future versions of DarkChem to graph convolution molecular representations to improve the ability to generate increasingly realistic, stable compounds.

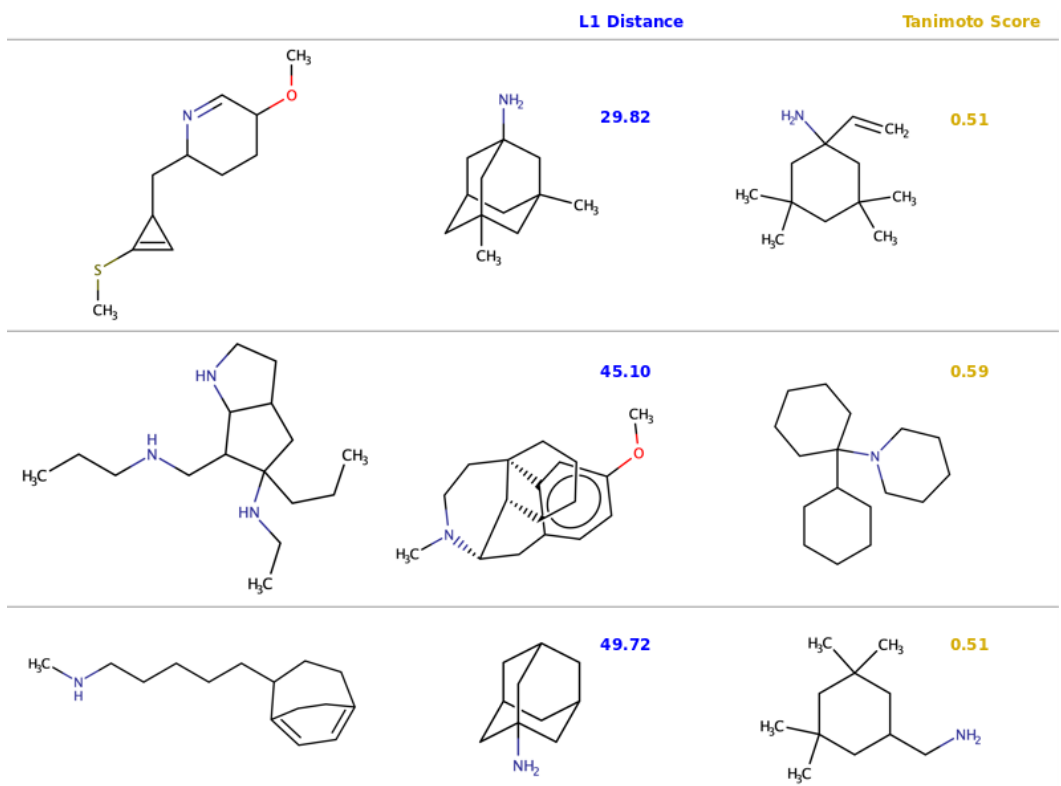

**Figure 5.** Finalist generated structures display a high degree of uniqueness compared to training set molecules. Representative finalist compounds, the three leftmost molecules, are juxtaposed with the most similar training set therapeutic by L1 distance and the most similar training set active by Tanimoto score.

Current State of AI for *De Novo* Molecular Design

AI is demonstrating promise as a tool in the drug design and discovery process. However, the field is in its infancy and is regularly described with either over-inflated hype about what it can deliver or harsh derision and dismissal. As such, many in the field are calling for a more measured approach coupled with the development of suitable guidelines for measuring performance of AI models to better understand current limitations and possibilities.[104] While sensationalist headlines would have one believe that we are nearing the point in which AI can effectively generate novel drugs with desirable *in vivo* properties for a target of interest, the reality is that current generative models have a considerable way to go before that claim can reasonably be made. In fact, it may not be possible for a single AI to autonomously design drugs; it could be that the best outcome will arise from a suite of AI models used in concert at various stages of the drug design and optimization process. Furthermore, the ideal model(s) might be dependent on the individual target rather than being universal.[105] There are many suggestions regarding specific intermediate goals that the AI drug design community should strive to achieve in an effort to establish standards by which generative models can be judged. One such suggested near-term goal is for generative models to demonstrate the effective shrinking of the vast search space used by traditional virtual and high-throughput screening methods.[106] However, the means by which to demonstrate success at such a task is up for debate. A large obstacle facing the AI drug design community is that there is no current "best practice" for assessing performance. It is widely agreed that at a minimum, the similarity of generated molecules to training molecules should be assessed. Other suggestions include comparing output to that produced by other drug design tools and bioisosteric replacement methods, assessment of activity at off-targets, and sensitivity analysis of how data quality and quantity affect output. To begin to fill this gap, some nascent benchmarking sets have recently been introduced.[8, 107] While the authors highlight the inherent difficult in creating objective quality metrics and note that many of the benchmarks are too easily solved by most DL-based *de novo* design models, the benchmarks offer a promising initial step in the development of robust quality assessment techniques.

CONCLUSION. In summary, we produced a library of NMDAR PCP site antagonists including known actives, known inactives, and decoys. We demonstrated the application of this library to the AI-assisted generation of structurally unique putative active compounds targeting a more complex site than previously seen that were then downselected using standard CADD techniques.

We identified current limitations of generative models in *de novo* design including the absence of receptor structural information during the training process and the lack of existing baselines for comparison. While the ability of AI-generated compounds to pass ligand- and structure-based filters for a complex target is promising, we advise that rigorous benchmarks are needed for more robust assessment of this technology.

METHODS.

NMDAR Antagonist Library

The NMDAR PCP site antagonist library was built by extracting from a comprehensive literature search for experimental $K_i$ and $IC_{50}$ values of ligands used in PCP site binding assays.[33, 36, 39-71, 73-88, 108-116] Ligands with $K_i$ values less than 100 µM were placed in the 'Actives' section of the library, and those 100 µM or greater were placed in the 'Inactives' section. The library contains 728 active and 87 inactive literature-verified compounds, and 297 PCP analogs and near-analogs without binding data scraped from the website [isomerdesign.com](isomerdesign.com). The html code from the page was parsed using the Python library BeautifulSoup[117] by visiting every compound stored under the tag 'arylcycloalkylamine' and extracting the molecular information. A set of thirty-nine actives from a single study[83] ('validation1' in Figure S1) with a wide range of PCP site activities were used to seed the generation of decoy molecules using the Database of Useful Decoys: Enhanced (DUD-E) database,[118] which were added to the library. All library compounds are represented by their canonical SMILES. Canonicalization was performed using OpenBabel (version 2.4.1, 2019, http://openbabel.org).[119] The complete library can be found in Supporting Information.

Ligand- and Structure-Based Analysis

Both 1- and 2-D physiocochemical properties were calculated for each of library compounds using ChemAxon's cxcalc (version 5.2.0, http://www.chemaxon.com). The properties were normalized and applied to the chemical property PCA space using the Python library scikit-learn,[120] in which PCA space was defined following the method outlined in the Colby et al. DarkChem publication.[121]

The library compounds were split into training and test sets (75%/25%) using stratified random sampling with ten-fold cross-validation to account for the imbalance between classes. An SVM binary classification model for predicting binding affinity was built and trained on the training set with scikit-learn, with performance assessed using recall, precision, f1 score, and area under the precision-recall curve (AUPR). The properties used in model building are enumerated in Table S1.

Structure-based analysis was performed using a 3D crystallographic structure of the NMDAR (PDB code 5UN1) downloaded from www.rcsb.com.[30] The receptor was prepared with Chimera[122] by removing the bound ligand MK-801 after creating a centroid to extract the center coordinates of the bound ligand. Next, the Chimera structure editing "dock prep" feature was run with the option to consider H-bonds, the AMBER ff14sb option to include charges for standard residues, and the AM1-BCC option to include charges for non-standard residues. The resulting NMDAR structure was saved for docking in PDB format after removing all hydrogens. Using this structure, the active, inactive, and decoy compounds from the library were run through an AutoDock Vina[123] docking simulation using PyRX[124], where the previously extracted centroid coordinates were used to define the search space in the protein. The top eight pose scores, or the maximum number found if below eight, were collected for each compound.

*De Novo* Design of Potential Channel Blockers

Putative structures were generated using DarkChem, a SMILES-based VAE implementation with coupled property predictor, with model parameters matching those of Colby et al.[26] The model was initialized with pretrained weights from a transfer learning configuration including ~55M mass-labeled compounds from PubChem;[100] ~700K mass- and computed collision cross section (CCS)- labeled compounds from the union of the Universal Natural Product Database,[125] Human Metabolome Database,[126] and Distributed Structure-searchable Toxicity[127] datasets; and ~500 mass- and experimental CCS- labeled compounds curated from the literature. This pretrained model was used to encode the set of 728 known NMDAR actives into a 128-dimension latent vector representation of molecular structure. The mean and variance of the actives was calculated for each dimension and used to define a 128-dimension random normal distribution from which 100K latent vectors were sampled. Each latent vector was decoded to the $k$ most probable structures using beamsearch (k=100), resulting in 10M putative structures. These were initially downselected by SMILES canonicalization and duplicate removal, checking for SMILES validity using *rdkit* (rdkit.org), and ensuring sampled latent vectors fell within the convex hull defined by the 728 known actives, reducing the putative set to 198,826. The convex hull was constructed from the first 8 principal components of the latent representation using the Quickhull[128] algorithm from the spatial module of SciPy.[129]

Downselection of Generated Compounds

The remaining structures were assessed for potential PCP site activity by being tested against the classification model. Passing compounds were then run through the same docking simulation as the library compounds. The compounds with docking scores in the range of known actives were assessed for synthesizability and filtered for desirable ADME properties using SwissADME,[130] for which the known therapeutic PCP site actives memantine, ketamine, dextromethorphan, and amantadine were used as a basis for comparison. A synthetic accessibility score of six or higher (on a scale of one to ten) was considered undesirable. Drug- and lead-likeness violations were only accepted if any of the known PCP site therapeutics contained the same violation. Predicted activity at any of the set of receptors checked by SwissADME was also only accepted if any known therapeutics demonstrated the same prediction. Remaining compounds were then filtered for desirable gastrointestinal absorption, blood-brain barrier permeability, and solubility. The passing molecules were further compared against the known therapeutics and filtered to those containing at least one hydrogen bond acceptor, no more than six heavy aromatic atoms, at least one nitrogen atom, and a consensus logP between two and four. A link to the Jupyter notebook containing the filtering script is included in Supporting Information. The twelve finalist compounds were cross-referenced against a chemical library comprised of a collection of thirty-one chemical databases and websites to determine their novelty.[125-126, 131-159] Predicted metabolites for each of the twelve finalist compounds were generated using BioTransformer[135] and are reported in Supporting Information.

Similarity Assessment of AI-Generated Compounds to Known Actives

Substructure assessment was conducted utilizing graph theory (i.e. subgraph isomorphism) where the molecular graph was defined by a set of nodes (i.e. atoms) and a set of connecting edges (i.e. bonds). Substructures ranging in size from one atom of a molecule to the entire molecule were found in two stages. In the first stage, only the paths consisting of a straight chain of nodes where each node is connected to every other in the molecule were found. Breadth-first search was used to find every possible path in the molecule by starting with an arbitrary atom in a molecular graph and exploring by degree, starting from all first-degree neighbor nodes and then moving to one-degree deeper nodes until all connecting nodes were visited. In the second stage, paths containing branches were searched using depth-first search by exploring each node branch to the deepest level from the starting node until no more connections were found. As a result, branches were mapped to the chain paths at possible positions. Obtained substructures were stored as canonical SMILES.

The distance between each finalist structure and active in the 128-dimension space was computed by L1 norm. The Tanimoto similarity between each of the twelve final structures and every library active was computed using RDKit. The similarity scores were evaluated to locate the nearest active by L1 norm and Tanimoto similarity to each finalist molecule.

ASSOCIATED CONTENT.

**Supporting Information**. NMDAR PCP site library, enumerated compounds at each stage of screening, additional figures and tables illustrating activity model assessment, PCA, docking numerics, and similarity assessment, ADME-based filtering script, QSAR SVM script, and predicted metabolites.

The following files are available free of charge.

NMDAR PCP Site Library (XLSX)

Supporting Figures and Tables (PDF)

Filtering Script (IPYNB)

QSAR SVM Script (IPYNB)

Predicted Metabolites of Finalists (XLSX)

AUTHOR INFORMATION

**Corresponding Author**

*Email: ryan.renslow@pnnl.gov

**Author Contributions**

The manuscript was written through contributions of all authors. All authors have given approval to the final version of the manuscript.

**Notes**

The authors declare no competing financial interests.

ACKNOWLEDGMENT

This research was supported by the Pacific Northwest National Laboratory (PNNL) Laboratory Directed Research and Development program. PNNL is operated for DOE by Battelle Memorial Institute under contract DE-AC05-76RL01830.

We thank Nathan Baker for his input on protein preparation for molecular docking simulations and Madison Blumer for providing BioTransformer assessment.

ABBREVIATIONS USED

ADME, absorption, distribution, metabolism and excretion; AI, artificial intelligence; CADD, computer-assisted drug discovery; CNS, central nervous system; DL, deep learning; NMDAR, *N*-methyl D-aspartate receptor; PCA, principal component analysis; PCP, phencyclidine; QSAR, quantitative structure-activity relationship; SMILES, Simplified Molecular Input Line Entry Specification; SVM, support vector machine; TMD, transmembrane domain; VAE, variational autoencoder.